\documentclass[conference]{IEEEtran}
\IEEEoverridecommandlockouts
\usepackage{cite}

\usepackage{amsmath,amssymb,amsfonts}

\usepackage{algorithmicx}

\usepackage{algpseudocode}
\usepackage{tikz}
\usetikzlibrary{fit}
\usetikzlibrary{tikzmark}
\usepackage{graphicx}
\usepackage{wrapfig}
\usepackage{textcomp}
\usepackage{pifont}
\usepackage{xcolor}
\usepackage{caption}
\usepackage{subcaption}
\usepackage{balance} 
\usepackage{xspace}
\usepackage[T1]{fontenc}
\usepackage[scaled=0.81]{beramono}
\usepackage{balance}

\usepackage{enumitem}  
\usepackage{listings}
\usepackage{url}
\usepackage[ruled,lined,linesnumbered,vlined]{algorithm2e}
\usepackage{multirow}
\usepackage{lscape}
\usepackage{longtable}
\usepackage{array} 
\usepackage{array}
\usepackage{tcolorbox}
\usepackage{color}
\usepackage[T1]{fontenc}
\usepackage[scaled=0.81]{beramono}
\usepackage{fancyvrb}
\usepackage{listings}
\usepackage{multicol}
\usepackage{lipsum}
\usepackage{booktabs}
\usepackage{listings}
\usepackage{color}
\usepackage{graphicx}
\usepackage{calc}
\usepackage{amsmath}
\usepackage{upgreek}
\usepackage{float}
\usepackage{mathrsfs}
\usepackage{amssymb}
\usepackage{threeparttable}
\usepackage[framemethod=tikz]{mdframed}
\usepackage{lipsum}
\usepackage{float}
\usepackage{amsthm}
\usepackage{pifont}

\usepackage{color}
\usepackage{wasysym}
\usepackage{colortbl}

\usepackage{perpage} 
\MakePerPage{footnote} 

\usepackage{listings, xcolor}
\usepackage{color}          

\definecolor{verylightgray}{rgb}{.97,.97,.97}

\lstdefinelanguage{Solidity}{
	keywords=[1]{anonymous, assembly, assert, balance, break, call, callcode, case, catch, class, constant, continue, contract, debugger, default, delegatecall, delete, do, else, emit, event, export, external, false, finally, for, function, gas, if, implements, import, in, indexed, instanceof, interface, internal, is, length, library, log0, log1, log2, log3, log4, memory, modifier, new, payable, pragma, private, protected, public, pure, push, require, return, returns, revert, selfdestruct, send, storage, struct, suicide, super, switch, then, this, throw, transfer, true, try, typeof, using, value, view, while, with, addmod, ecrecover, keccak256, mulmod, ripemd160, sha256, sha3}, 
	keywordstyle=[1]\color{blue}\bfseries,
	keywords=[2]{address, bool, byte, bytes, bytes1, bytes2, bytes3, bytes4, bytes5, bytes6, bytes7, bytes8, bytes9, bytes10, bytes11, bytes12, bytes13, bytes14, bytes15, bytes16, bytes17, bytes18, bytes19, bytes20, bytes21, bytes22, bytes23, bytes24, bytes25, bytes26, bytes27, bytes28, bytes29, bytes30, bytes31, bytes32, enum, int, int8, int16, int24, int32, int40, int48, int56, int64, int72, int80, int88, int96, int104, int112, int120, int128, int136, int144, int152, int160, int168, int176, int184, int192, int200, int208, int216, int224, int232, int240, int248, int256, mapping, string, uint, uint8, uint16, uint24, uint32, uint40, uint48, uint56, uint64, uint72, uint80, uint88, uint96, uint104, uint112, uint120, uint128, uint136, uint144, uint152, uint160, uint168, uint176, uint184, uint192, uint200, uint208, uint216, uint224, uint232, uint240, uint248, uint256, var, void, ether, finney, szabo, wei, days, hours, minutes, seconds, weeks, years},	
	keywordstyle=[2]\color{teal}\bfseries,
	keywords=[3]{block, blockhash, coinbase, difficulty, gaslimit, number, timestamp, msg, data, gas, sender, sig, value, now, tx, gasprice, origin},	
	keywordstyle=[3]\color{violet}\bfseries,
	identifierstyle=\color{black},
	sensitive=false,
	comment=[l]{//},
	morecomment=[s]{/*}{*/},
	commentstyle=\color{gray}\ttfamily,
	stringstyle=\color{red}\ttfamily,
	morestring=[b]',
	morestring=[b]"
}

\definecolor{darkgreen}{rgb}{0.0, 0.5, 0.13}
\definecolor{C-gray}{gray}{0.85}
\definecolor{B-gray}{gray}{0.65}
\definecolor{A-gray}{gray}{0.95}

\newcommand{\tool}{\hbox{\textsc{UAT20}}\xspace}

\newcommand{\ethereum}{\hbox{Ethereum}\xspace}

\newcommand{\myparagraph}[1]{\vspace*{0.14cm}\noindent\textbf{\emph{#1.}}\quad}
\theoremstyle{definition}
\newtheorem{thm}{Theorem}
\newtheorem{defn}[thm]{Definition}

\newcommand{\eg}{\hbox{\emph{e.g.}}\xspace}
\newcommand{\ie}{\hbox{\emph{i.e.}}\xspace}

\begin{document}
\lstset{%
	escapeinside={(*@}{@*)}
}

\title{\textsc{UAT20}: Unifying Liquidity Across Rollups}



\author{
	\IEEEauthorblockN{Yue Li \quad Han Liu}
	\IEEEauthorblockA{\textit{UAT Team} \\
	} 
}

\maketitle

Ethereum has been a cornerstone of the decentralized ecosystem, with rollup-based scaling solutions like Arbitrum and Optimism significantly expanding its capabilities. These rollups enhance scalability and foster innovation, but their rapid proliferation has introduced \emph{liquidity fragmentation}. Specifically, tokens distributed on multiple rollups fragment the liquidity of users, complicating participation in trading and lending activities bound by minimum liquidity thresholds.

This paper proposes UAT20, a universal abstract token standard, to address liquidity fragmentation across rollups. Leveraging Conflict-free Replicated Data Types (CRDTs), UAT20 ensures consistent states across multiple rollups. We introduce a two-phase commit protocol to resolve transaction conflicts, enabling seamless and secure liquidity unification. 
Finally, our empirical analysis demonstrated the necessity and effectiveness of UAT20 in mitigating liquidity fragmentation within Rollups.


\section{Introduction}
\label{sec:intro}
Ethereum, as one of the earliest smart contract platforms, continues to play a pivotal role in the evolving multi-chain ecosystem\cite{Wood2014Ethereum}. The introduction of rollup-based scaling solutions, such as Arbitrum and Optimism, paved the way for designing Ethereum Layer\cite{kalodner2018arbitrum, donno2022optimistic, scroll,zksync}. Furthermore, the modular and replicable nature of underlying technology has made it possible to create new rollups quickly.
For example, over 50 chains have been launched based on the OP Stack's Superchain\cite{superchain}. These rollups enhance Ethereum's scalability and expand the development landscape, further enriching Ethereum's ecosystem. However, this rapid proliferation of rollups has also introduced issues, with \emph{liquidity fragmentation} emerging as a significant concern. Specifically, 

\myparagraph{Liquidity Fragmentation} refers to the fragmentation of available liquidity for a particular asset across multiple chains. This fragmentation can lead to i) inefficiencies in trading and lending activities in a specific chain; ii) asset price inefficiencies and increased volatility, even small trades can significantly impact asset prices\cite{lehar2023liquidity}.
In \ethereum's rollup ecosystem, asset issuers tend to deploy their token on multiple rollups to maximize economic opportunities and attract a broader user base. Moreover, cross-chain bridges also enable users to transfer their assets among rollups\cite{whinfrey2021hop, xie2022zkbridge, lan2021horizon,zetachain,chainflip,Interlay}.
Thus, the liquidity fragmentation has emerged as a significant challenge within Ethereum's rollup ecosystem.

Currently, users often rely on cross-chain bridges to manually transfer their liquidity, which not only increases the complexity and cost of participating in DeFi but also introduces additional security risks\cite{lehar2023liquidity, augusto2024sok}.
Moreover, a range of DEX aggregators\cite{1inch,swoopexchange,matcha} has been proposed to provide a seamless trading experience. The underlying technologies enabling these aggregators typically involve cross-chain bridges and token pre-authorization mechanisms. However, these works have the following limitations: (i) only limited transaction scenarios are supported, such as cross-chain token swaps; (ii) a third-party authorization is required, which also introduces additional security risks.

\myparagraph{The Insight of \tool} 
The goal of this paper is to mitigate liquidity fragmentation across rollups without introducing the aforementioned limitations. We propose leveraging Conflict-free Replicated Data Types (CRDTs)—abstract data types that converge to the same state in distributed environments\cite{preguicca2018conflict,enes2019efficient}—as a promising solution. Specifically, we introduce a universal abstract token standard, UAT20, which unifies a user's ERC20 tokens across multiple rollups. The CRDT forms the foundation of the UAT20 data structure, ensuring consistency across multiple rollups. To enable seamless operation of UAT20 tokens on any rollup, we design a two-phase commit protocol to resolve conflicts arising from token transactions across rollups. This protocol ensures that 
unified liquidity is updated consistently across all rollups, preserving convergence and reliability within Rollups.


\myparagraph{Main Contributions} 
We summarize the main contributions of this paper as below.

\begin{itemize}[leftmargin=*]


\item We proposed a novel token standard called universal abstract token (UAT20) to solve liquidity fragmentation in Rollups without introducing security risks.

\item We present a two-phase commit protocol for UAT20, which combines token operations on different Rollups while handling conflicts between them in a uniform way.

\item We empirically studied historical transactions from three rollups to quantify the generality of liquidity fragmentation in Ethereum's rollup ecosystem, reflecting the necessity and effectiveness of UAT20.



\end{itemize}


\section{The \tool Protocol}
\label{sec:protocol}

\begin{figure*}[h!]
\centering
\includegraphics[width=.7\linewidth]{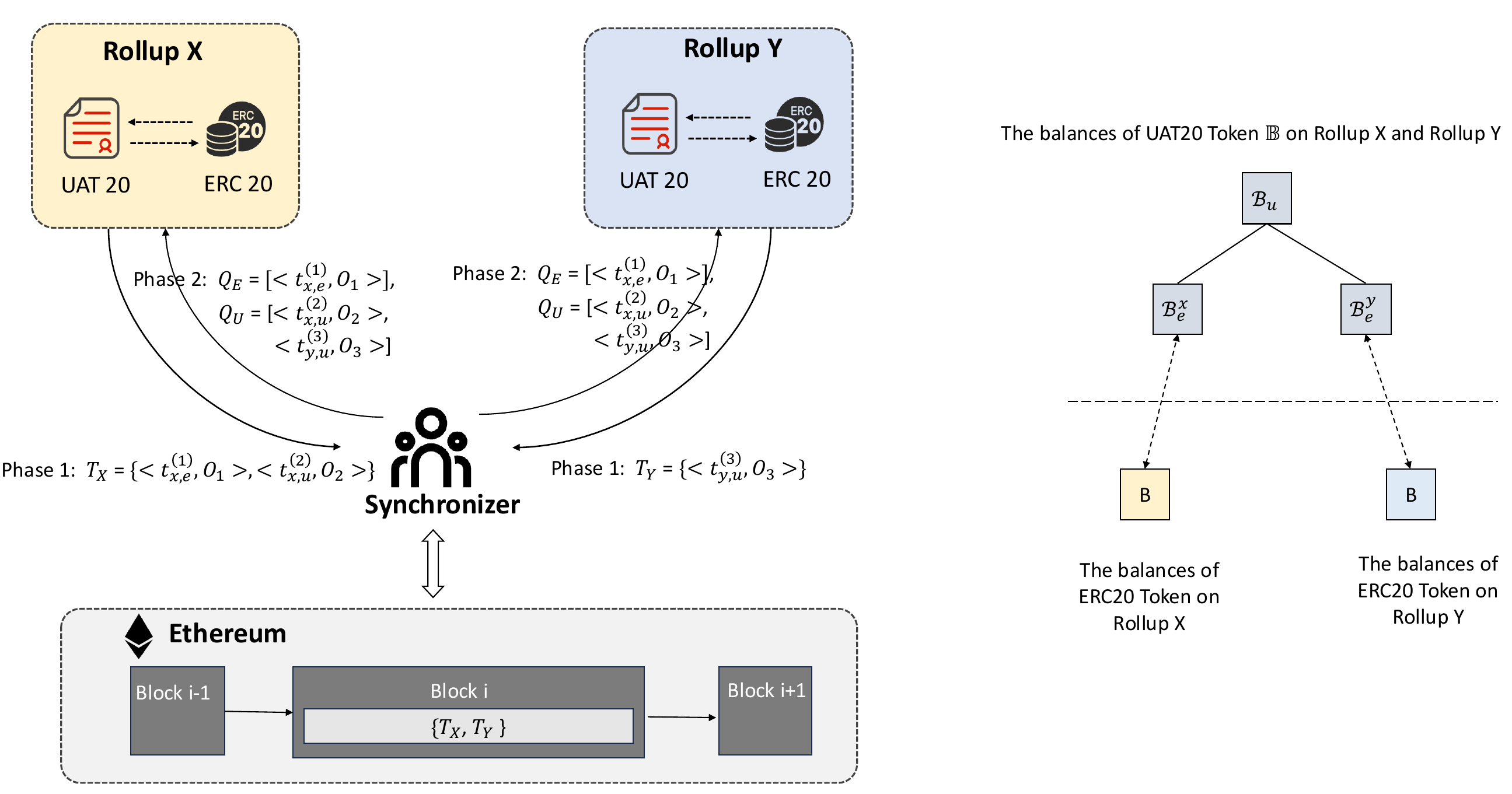}
\caption{\label{fig:arch} The Workflow of \tool protocol}
\end{figure*}
 

\subsection{System Model}

\myparagraph{User, Ethereum, Rollups} UAT20 is designed for unifying liquidity among Rollups $\{\mathcal{R}_1, \mathcal{R}_2, \cdots, \mathcal{R}_n\}$. Ethereum $\mathcal{E}$ servers as the base layer where all Rollups ultimately submit their batched transactions for final confirmation. Each Rollup $\mathcal{R}_i$, whether based on Optimistic or Zero-Knowledge (ZK) technologies, operates independently by maintaining its own state and processing user transactions to update this state. These rollups rely on Ethereum (L1) for transaction finality and confirmation, ensuring security and consensus.
Users $\mathcal{U}$ are allowed to send transactions in both Ethereum (L1) and Rollups (L2).

\myparagraph{ERC20 Token ($\Sigma_{20}$), UAT20 Token ($\Upsilon_{20}$)} Two types of tokens contract is deployed in Rollups: ERC20 Token ($\Sigma_{20}$) and UAT20 Token ($\Upsilon_{20}$).
$\Sigma_{20}$ follows the mature ERC-20 standard, preserving user assets on the current chain and enabling token transfers within the same chain. However, user tokens remain fragmented across different instances of the same $\Sigma_{20}$ deployed on multiple chains.
As a universal abstract token built on top of E-Token, $\Upsilon_{20}$
  manages all of a user's ERC20 tokens across multiple chains, enabling seamless asset transfers on any chain while ensuring consistency across them. $\Upsilon_{20}$ leverages a Conflict-Free Replicated Data Type (CRDT) design, allowing each replica (\ie, Rollup) to independently handle token transfers. To maintain a unified state across all rollups, $\Upsilon_{20}$ employs merge functions that reconcile concurrent updates from different replicas (\ie, Rollup), ensuring consistency and reliability in multiple rollups.





\myparagraph{Synchronizer} A synchronizer is a critical component that enables rollups to broadcast their token transaction (e.g., the transaction of $\Sigma_{20}$ or $\Upsilon_{20}$) to Ethereum, allowing other rollups to access these transactions. Once Ethereum organizes transactions from multiple rollups into a new block, the Synchronizer ensures that the sorted 
transactions are synchronized across all rollups. In summary, the synchronizer operates with two functions:

\noindent (i) $\mathtt{Broadcast}(T_i, \mathcal{R}_i)$: facilitates the broadcast of token transfer transactions $T_i$ from the rollup $\mathcal{R}_i$;

\noindent (ii) $\mathtt{Sync}(\mathbb{T}_{O}, \mathcal{R}_i)$: takes the sorted transactions $\mathbb{T}_{O}$ from Ethereum and synchronizes them to the target rollup $\mathcal{R}_i$.



\myparagraph{UAT20 Requirements} We define a set of requirements that UAT20 must satisfy to effectively address liquidity fragmentation across rollups. First, \emph{\textbf{compatibility}}: UAT20 should minimally impact both the original rollup design and the underlying ERC20 token ($\Sigma_{20}$) design within the Ethereum ecosystem. The second requirement is \emph{\textbf{eventual consistency}}: UAT20 must successfully commit all valid token transactions. The state of $\Upsilon_{20}$ should eventually converge to a consistent state across all rollups, ensuring that all valid updates to tokens are preserved.




\subsection{Workflow}

As shown in Figure\ref{fig:arch} , Rollup \texttt{X} ($\mathcal{R}_X$) and Rollup \texttt{Y} ($\mathcal{R}_Y$) is deployed with two types of token: ERC20 token $\Sigma_{20}$ and UAT20 Token $\Upsilon_{20}$. Within each rollup, the $\Upsilon_{20}$ is designed based on the CRDT framework as a replica, managing and reflecting users' $\Sigma_{20}$ across all rollups. 
Users can perform transactions to transfer or approve either $\Upsilon_{20}$ or $\Sigma_{20}$ on any rollup.
These transactions are periodically broadcast to Layer 1 (\ethereum). \ethereum serves as the coordinator, responsible for ordering token transactions from multiple rollups within its blocks. 
This transaction sequence allows each rollup to merge transactions and resolve conflicts arising from concurrent transactions in a consistent manner, ensuring that all rollups converge to the same state for $\Upsilon_{20}$. Specifically, the update of $\Upsilon_{20}$'s state follows a two-phase \textbf{Execute-Commit} protocol.


\myparagraph{Phase1: Execution Phase}
 Users first submit transaction $t$ to invoke $\Sigma_{20}$ or $\Upsilon_{20}$ in each rollup, resulting in a set of operations to be committed in the commit phase. After these transactions are executed,  the \emph{synchronizer} forwards these transactions and their operations to Ethereum for ordering. As shown in Figure~\ref{fig:arch}, the user $\textbf{C}$ initates two transactions on $\mathcal{R}_X$: $T_X = \{t^{(1)}_{x, e}, t^{(2)}_{x,u}\}$.
 The first transaction $t^{(1)}_{x, e}$ invokes $\Sigma_{20}$ to transfer ERC20 tokens within $\mathcal{R}_X$, while $t^{(2)}_{x,u}$ invokes $\Upsilon_{20}$ to enable the transfer of UAT20 tokens across rollups. 
 The execution of these transactions generates a set of operations, \eg, the operation set $O_1$ for $t^{(1)}_{x, e}$, which are derived using the CRDT approach and serve to update the state of $\Upsilon_{20}$ across all replicas (rollups). We will detail the execution of these transactions and their operations in \S\ref{sec:design}.
Finally, transactions and their operations are periodically collected by the \emph{synchronizer} and submitted to Ethereum for ordering.
 

\myparagraph{Phase2: Commit Phase}
The \emph{synchronizer} then synchronizes the ordered transactions sequence $\mathbb{T}_{O}$ from \ethereum. Note that the $\mathbb{T}_{O}$ specifies the \emph{arbitration order} for committing.
Specifically, $\mathbb{T}_{O}$ consists of two queues $[\mathcal{T}_E, \mathcal{T}_U]$ where $\mathcal{T}_E$ contains transactions calling $\Sigma_{20}$ from multiple rollups, while the latter $\mathcal{T}_U$ contains transactions calling $\Upsilon_{20}$ from multiple rollups.
The \emph{arbitration order} specifies that: (i) Transactions in $\mathcal{T}_E$ have higher processing priority than those in $\mathcal{T}_U$; 
(ii) Within each queue, transactions are processed in the order, following the sequence of the queue.
For each transaction's set of operations, the function $\mathtt{commit}$ of $\Upsilon_{20}$ will be to apply these operations to their states and update these state according to the update strategy (refer to \S\ref{sec:design}).
In this way, conflicts between non-commuting concurrent operations from two CRDTs are resolved based on this arbitration order.

\myparagraph{Example} Figure~\ref{fig:arch} shows an example of token transfer in UAT20.
Consider user \textbf{C} has 100 UAT20 token (\ie, $\mathcal{B}_u[\mathbf{C}] = 100$), which is composed of 40 ERC20 tokens on $\mathcal{R}_X$ (\ie, $\mathcal{B}^x_e[\mathbf{C}] = 40$) and 60 ERC20 tokens on $\mathcal{R}_X$ (\ie, $\mathcal{B}^y_e[\mathbf{C}] = 60$). During a block epoch, \textbf{C} initiates three transactions: $t^{(1)}_{x,e}, t^{(2)}_{x,e}$ on $\mathcal{R}_X$ and $t^{(3)}_{y,u}$
on $\mathcal{R}_Y$. Specifically:

(i) the $t^{(1)}_{x,e} = \Sigma_{20}.\mathtt{transfer}(\mathbf{C}\rightarrow \mathbf{A},10)$ transfers 10 ERC20 token from user \textbf{C} to user \textbf{A} via the $\Sigma_{20}$ on $\mathcal{R}_X$; 

(ii) the $t^{(2)}_{x,u} = \Upsilon_{20}.\mathtt{transfer}(\mathbf{C}\rightarrow \mathbf{A},50)$ transfers 50 UAT20 tokens from $\mathbf{C}$ to $\mathbf{A}$ via the $\Upsilon_{20}$ on $\mathcal{R}_X$;

(iii) the $t^{(3)}_{y,u} = \Upsilon_{20}.\mathtt{transfer}(\mathbf{C}\rightarrow \mathbf{A},90)$ transfers 90 UAT20 tokens from user \textbf{C} to user \textbf{A} via the $\Upsilon_{20}$ on $\mathcal{R}_Y$.

For each transaction, an operation set $O$ is generated to describe the token transfers within that transaction. Once executed, the synchronizer periodically submits these transactions along with their operation sets to Ethereum for inclusion in a new block. Assume Ethereum finalizes the following order: $[\langle t^{(1)}_{x,e}, O_1 \rangle, \langle t^{(2)}_{x,u}, O_2 \rangle, \langle t^{(3)}_{y,u}, O_3 \rangle]$.
The synchronizer facilitates the synchronization of token transactions across rollups by interacting with the $\Upsilon_{20}$ on each $\mathcal{R}$. According the type of transferred token, these transactions are divided into two queues $\mathcal{T}_E = [\langle t^{(1)}_{x,e}, O_1 \rangle]$ and $\mathcal{T}_U = [\langle t^{(2)}_{x,u}, O_2 \rangle, \langle t^{(3)}_{y,u}, O_3 \rangle]$.
The $\Upsilon_{20}$ processes these transactions sequentially, starting with $\mathcal{T}_E$ and followed by $\mathcal{T}_U$. During this process, a conflict arises between $t^{(2)}_{x,u}$ and $t^{(3)}_{y,u}$. Specifically, the $\mathcal{B}_u[\mathbf{C}]$ is insufficient to fulfill the above two transactions (90 tokens available, but $t^{(3)}_{y,u}$ requires 90 tokens and $t^{(2)}_{x,u}$ requires 50 tokens). To resolve this conflict, the UAT20 adheres to the order of the $\mathcal{T}_U$. Consequently, $t^{(2)}_{x,u}$ succeeds as it appears earlier in the queue, while $t^{(3)}_{y,u}$ fails due to insufficient balance.


\subsection{UAT20 Token and ERC20 Token}
\label{sec:design}

In this section, we present the design of ERC20 Token $\Sigma_{20}$ and UAT20 Token $\Upsilon_{20}$ in detail.


\begin{defn}[UAT20 Token $\Upsilon_{20}$] The $\Upsilon_{20}$ is a tuple $(\mathbb{B}, \mathcal{F}_o, \mathcal{F}_c)$ where:
\begin{itemize}[leftmargin=*]
\item $\mathbb{B}$ is represents the aggregated token state across $n$ Rollups $\{ \mathcal{B}_u, \mathcal{B}^1_e, \mathcal{B}^2_e, \cdots, \mathcal{B}^n_e \}$ where $\mathcal{B}_u$ is defined as the sum of all balances of $\Sigma_{20}$ maintained within the individual Rollups, $\mathcal{B}^i_e$ is a mapping variable to the balance of ERC20 token held by each user on $\mathcal{R}_i$. For a user $\mathbf{C}$, 
$
\mathcal{B}_u[\mathbf{C}] = \sum_{i=1}^{n} \mathcal{B}^i_e[\mathbf{C}]$. 

\item $\mathcal{F}_o: \{\mathtt{transfer}, \mathtt{transferFrom}, \mathtt{Approve}\}$ denotes a collection of functions enabling users to transfer UAT20 token. Once the validity checks (\eg, permission check) are passed, a set of operations $O= [op_1, \cdots, op_n]$ is generated.
\item $\mathcal{F}_c: \{\mathtt{commitE}, \mathtt{commitU}\}$ empowers the synchronizer to submit token transactions from various Rollups within a block epoch during the commit phase.  In these functions, these token operations are finalized and synchronized across replicas (Rollups) using CRDT properties to maintain eventual consistency.
Specifically, \texttt{commitE} processes transactions and their operations from the $\mathcal{T}_E$, while \texttt{commitU} handles transactions and their operations from the $\mathcal{T}_U$. 
Since ERC20 transactions in $\mathcal{T}_E$ only involve updates to token within a single rollup, they are conflict-free. As such, \texttt{commitE} is 
responsible for updating the global balances $\mathcal{B}_u$ and the balances of involved rollup $\mathcal{B}^i_e$ based on the operations specified in each transaction.
Unlike \texttt{commitE}, conflicting transactions in $\mathcal{T}_U$ must be processed in \emph{arbitration order} to ensure eventual consistency.
Moreover, since operations of each UAT20 transaction only specify the update of the $\mathcal{B}_u$, \texttt{commitE} must resolve these global updates into specific rollup balances ($\mathcal{B}^1_e, \mathcal{B}^2_e, \cdots, \mathcal{B}^n_e$) based on the user's predefined priority strategy (which relies on Conflict-free Replicated Data Types (CRDTs) to ensure consistent agreement across all rollups), \eg,  prioritizing the deduction of tokens from $\mathcal{R}_X$ in Figure \ref{fig:arch}. 
\end{itemize}


\end{defn}

\begin{defn}[ERC20 Token $\Sigma_{20}$] $\Sigma_{20}$ enhances the ERC20 standard by incorporating functionality that supports the UAT20 protocol, facilitating advanced global token management. The update of $\Sigma_{20}$'s state will be forwarded to $\Upsilon_{20}$ by calling the function in $\mathcal{F}_o$, which produces a set of token operations to be committed.
Overall, E-Token is defined as a tuple $(B, F)$, where:

\begin{itemize}[leftmargin=*]
    \item $B$ is a mapping variable that records the balance of each user; 
    \item $F = (\mathtt{transfer}, \mathtt{transferFrom}, \mathtt{Approve})$ denotes a collection of functions enabling users to transfer UAT20 token.
    $\mathcal{F}_e$ is invoked at the end of these functions to produce operations for updating $\mathcal{B}_u$ and $\mathcal{B}^i_e$ ($i$ is the id of the current rollup and $\mathcal{B}^i_e = B$); Beside, $F$ can also be invoked by the $\mathcal{F}_c$ to update $B$ during the commit phase.
\end{itemize}


\end{defn}

We still use the example in Figure \ref{fig:arch} to explain the workflow of $\Sigma_{20}$ and $\Upsilon_{20}$.

\myparagraph{The transfer of $\Sigma_{20}$ on $\mathcal{R}_X$}
(\emph{Execution Phase.}) User \textbf{C} initiates the transaction $t^{(1)}_{x,e}$ by calling the function \texttt{tranfer} of $\Sigma_{20}$. During the execution, this transaction first check the validity of $t^{(1)}_{x,e}$ to ensure that the sender has sufficient E-Token on $\mathcal{R}_X$ and validate the transaction metadata(\ie, \texttt{msg.sender} or signature).
After that, the balances of $\Sigma_{20}$ for caller \textbf{C} and recipient \textbf{A} ($B[\mathbf{C}]$ and $B[\mathbf{A}]$)are updated accordingly.
Given that balances of $\Upsilon_{20}$ ($\mathbb{B}$) are unified across rollups, $t^{(1)}_{x,e}$ also affects both the Rollup-specific balances ($\mathcal{B}^x_e$) and the aggregated balances ($\mathcal{B}_u$). 
Thus, the output operations of $t^{(1)}_{x,e}$ is 
$O_1=\{\mathtt{SUB}(\mathcal{B}_u[\mathbf{C}],10), \mathtt{ADD}(\mathcal{B}_u[\mathbf{A}],10),\mathtt{SUB}(\mathcal{B}^i_e[\mathbf{C}],10),\mathtt{ADD}(\mathcal{B}^i_e[\mathbf{A}]$
$,10)\}$
(\emph{Commit Phase.}) Since $t^{(1)}_{x,e}$ is an ERC20 transfer transaction, there are no conflicts with other transaction at this epoch. With operations in $O_1$, $\Upsilon_{20}$ on each rollup updates both $\mathcal{B}_u$ and $\mathcal{B}^i_e$ balances for \textbf{A} and \textbf{C}, maintaining consistency across multiple rollups.

\myparagraph{The transfer of $\Upsilon_{20}$ on Rollup X}
(\emph{Execution Phase.}) User \textbf{C} initiates a transaction $t^{(2)}_{x,u}$ on $\mathcal{R}_x$ to transfer 50 UAT20 token to \textbf{A} by calling $\Upsilon_{20}.\mathtt{transfer}$.
This transaction first verifies transaction metadata (\ie, \texttt{msg.sender} or signatures), to confirm authenticity.
Then, a set of operations will be generated 
$O_2=\{\mathtt{SUB}(\mathcal{B}_u[\mathbf{C}],50), \mathtt{ADD}(\mathcal{B}_u[\mathbf{A}],50)\}$
During the execution phase, no actual balances are updated. Instead, the $\upsilon_{20}$ produces these operations as pending actions to be processed in the subsequent commit phase. 
(\emph{Commit Phase.}) $\mathcal{\Upsilon}_{20}$ processes operations in $O_2$, \ie, update $\mathcal{B}_u[\mathbf{C}]$ and $\mathcal{B}_u[\mathbf{A}]$. Notice that the $O_2$ does not explicitly specify how to update rollup-specific balances, \eg, $\mathcal{B}^x_e$ and $\mathcal{B}^y_e$.
 Instead, $\Upsilon_{20}$ relies on the user's predefined policy, which has been agreed upon across $\mathcal{R}_X$ and $\mathcal{R}_Y$, to determine the sequence and priority for balances deductions across rollups. In this case, $\mathbf{A}$’s policy prioritizes deducting tokens from $\mathcal{R}_X$ before $\mathcal{R}_Y$.
 Thus, 40 tokens are deducted in $\mathcal{B}^x_e[\mathbf{C}]$, depleting it from 40 to 0. Then, the remaining 10 tokens are deducted from $\mathcal{B}^y_e[\mathbf{C}]$, reducing it from 60 to 50. Simultaneously, $\mathcal{B}^x_e[\mathbf{A}]$ is increased by 40 and $\mathcal{B}^y_e[\mathbf{A}]$ is increased by 40.
 Additionally, these updates to the rollup-specific balances $\mathcal{B}^x_e$ and $\mathcal{B}^y_e$ will be synchronized to the balance of $\Sigma_{20}$ ($B$) on $\mathcal{R}_X$ and $\mathcal{R}_Y$ by calling $\Sigma_{20}.\mathtt{transfer}$.



\section{Empirical Study}
\label{sec:discussion}
This study seeks to empirically investigate the extent of liquidity fragmentation by analyzing token distribution and user behavior in Ethereum rollups so as to gain insight into the severity and impact of fragmentation. 

\myparagraph{Data Collection and Methodology}  We leverage the Google BigQuery platform \cite{google_bigquery} to access Arbitrum dataset and Optimism dataset. ZKsync is not included in Google BigQuery public datasets, thus we  downloaded the data from ZKSync via its Block Explorer API\cite{zksyncapi}. Overall, We gather state and historical transactions for 11 popular ERC20 tokens deployed on these three rollups: \textbf{USDT}, \textbf{USDC}, \textbf{wBTC}, \textbf{wETH}, \textbf{DAI}, \textbf{LINK}, \textbf{ZRO}, \textbf{HOP}, \textbf{AAVE}, \textbf{SYN} and \textbf{UNI}. Using SQL queries and Python scripts, we analyze token distribution, focusing on cases where users hold the same token across both rollups. Historical transaction data related to token-bridging is also examined to measure liquidity aggregation activity from January 1, 2024 to December 20, 2024.
A liquidity unification behavior is identified as follows:
a user executes a token transfer-out transaction on one rollup, following one or more token transfer-in transactions. Additionally, the user has carried out one or more token transfer-out transactions on other rollups in the two hours before.

\myparagraph{The result of liquidity fragmentation} The results indicate that 4,142,268 accounts hold the same token across multiple rollups. Among these, 1,096,997 accounts hold USDT on two rollups, while 145,193 accounts hold wETH on two rollups. Additionally, a total of 1,952,503 addresses hold the same token across three rollups, further highlighting the significant fragmentation of liquidity within the ecosystem.

This fragmentation reflects the diverse engagement of users with multiple rollups, which introduces significant complexity in liquidity management. These insights highlight the growing need for UAT20 to streamline user experience and unify liquidity.



\begin{figure}[h!]
\centering
\includegraphics[width=.8\linewidth]{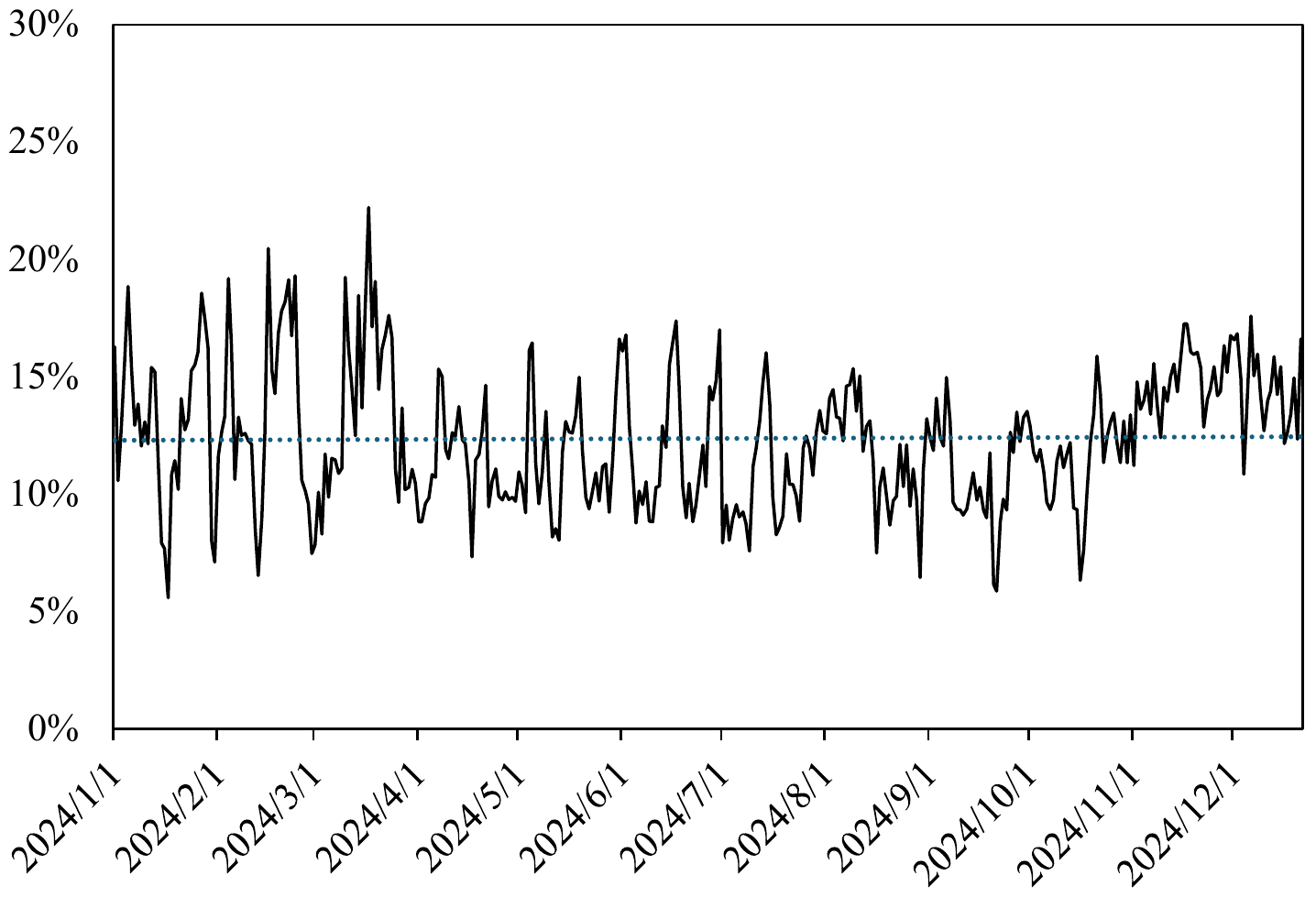}
\caption{\label{fig:r2} The result of liquidity unification}
\end{figure}

\myparagraph{The result of liquidity unification}
As shown in Figure \ref{fig:r2}, we analyzed transactions of 11 ERC20 tokens across three rollups to identify those related to liquidity unification. Among these, approximately 12.37\% of the transactions involved liquidity aggregation, with a daily peak reaching 22.21\%. 

This finding highlights that liquidity aggregation transactions constitute a significant portion of the overall token activity, resulting in notable resource consumption (\eg, gas consumption). Adopting the UAT20 offers a solution by unifying liquidity, thereby reducing the reliance on such transactions and enhancing overall efficiency.

\section{Conclusion}
\label{sec:conclusion}

This paper addresses the pressing issue of liquidity fragmentation within Ethereum’s rollups by introducing UAT20, a universal abstract token standard. By leveraging Conflict-free Replicated Data Types (CRDTs), UAT20 ensures consistent and unified token states across multiple rollups, eliminating the inefficiencies and complexities caused by fragmented liquidity. The proposed two-phase commit protocol resolves conflicts arising from UAT20 transactions, ensuring \emph{eventual consistency}. Through an empirical analysis of liquidity fragmentation, this work provides a robust solution that enhances liquidity utilization, simplifies user interactions, and strengthens the foundation of Ethereum’s multi-rollup ecosystem, paving the way for a more seamless and efficient decentralized financial landscape.

%

%
%
%
%

\bibliographystyle{IEEEtran}
\bibliography{ethereum,pl} 

\begin{thebibliography}{10}
\providecommand{\url}[1]{#1}
\csname url@samestyle\endcsname
\providecommand{\newblock}{\relax}
\providecommand{\bibinfo}[2]{#2}
\providecommand{\BIBentrySTDinterwordspacing}{\spaceskip=0pt\relax}
\providecommand{\BIBentryALTinterwordstretchfactor}{4}
\providecommand{\BIBentryALTinterwordspacing}{\spaceskip=\fontdimen2\font plus
\BIBentryALTinterwordstretchfactor\fontdimen3\font minus \fontdimen4\font\relax}
\providecommand{\BIBforeignlanguage}[2]{{%
\expandafter\ifx\csname l@#1\endcsname\relax
\typeout{** WARNING: IEEEtran.bst: No hyphenation pattern has been}%
\typeout{** loaded for the language `#1'. Using the pattern for}%
\typeout{** the default language instead.}%
\else
\language=\csname l@#1\endcsname
\fi
#2}}
\providecommand{\BIBdecl}{\relax}
\BIBdecl

\bibitem{Wood2014Ethereum}
G.~Wood, ``Ethereum: A secure decentralised generalised transaction ledger,'' \emph{Ethereum Project Yellow Paper}, vol. 151, 2014.

\bibitem{kalodner2018arbitrum}
H.~Kalodner, S.~Goldfeder, X.~Chen, S.~M. Weinberg, and E.~W. Felten, ``Arbitrum: Scalable, private smart contracts,'' in \emph{27th USENIX Security Symposium (USENIX Security 18)}, 2018, pp. 1353--1370.

\bibitem{donno2022optimistic}
L.~Donno, ``Optimistic and validity rollups: Analysis and comparison between optimism and starknet,'' \emph{arXiv preprint arXiv:2210.16610}, 2022.

\bibitem{scroll}
``Scroll is the leading zero-knowledge rollup. scaling ethereum for good.'' \url{https://scroll.io/}, 2022.

\bibitem{zksync}
``Zksync is an ever expanding verifiable blockchain network, secured by math,'' \url{https://zksync.io/}, 2022.

\bibitem{superchain}
``The superchain ecosystem platform accelerates the adoption and development of the op stack.'' \url{https://www.superchain.eco/}, 2023.

\bibitem{lehar2023liquidity}
A.~Lehar, C.~Parlour, and M.~Zoican, ``Liquidity fragmentation on decentralized exchanges,'' \emph{arXiv preprint arXiv:2307.13772}, 2023.

\bibitem{whinfrey2021hop}
C.~Whinfrey, ``Hop: Send tokens across rollups,'' 2021.

\bibitem{xie2022zkbridge}
T.~Xie, J.~Zhang, Z.~Cheng, F.~Zhang, Y.~Zhang, Y.~Jia, D.~Boneh, and D.~Song, ``zkbridge: Trustless cross-chain bridges made practical,'' in \emph{Proceedings of the 2022 ACM SIGSAC Conference on Computer and Communications Security}, 2022, pp. 3003--3017.

\bibitem{lan2021horizon}
R.~Lan, G.~Upadhyaya, S.~Tse, and M.~Zamani, ``Horizon: A gas-efficient, trustless bridge for cross-chain transactions,'' \emph{arXiv preprint arXiv:2101.06000}, 2021.

\bibitem{zetachain}
\BIBentryALTinterwordspacing
``Zetachain: the future of multichain.'' 2022. [Online]. Available: \url{https://www.zetachain.com/}
\BIBentrySTDinterwordspacing

\bibitem{chainflip}
\BIBentryALTinterwordspacing
``Native cross chain swaps.'' 2021. [Online]. Available: \url{https://chainflip.io/}
\BIBentrySTDinterwordspacing

\bibitem{Interlay}
\BIBentryALTinterwordspacing
``Interlay:use your bitcoin. anywhere.'' 2021. [Online]. Available: \url{https://interlay.io/}
\BIBentrySTDinterwordspacing

\bibitem{augusto2024sok}
A.~Augusto, R.~Belchior, M.~Correia, A.~Vasconcelos, L.~Zhang, and T.~Hardjono, ``Sok: Security and privacy of blockchain interoperability,'' in \emph{2024 IEEE Symposium on Security and Privacy (SP)}.\hskip 1em plus 0.5em minus 0.4em\relax IEEE, 2024, pp. 3840--3865.

\bibitem{1inch}
``1inch network offers a defi ecosystem with products like 1inch dapp, wallet, developer portal, portfolio, and fusion for secure web3 operations.'' \url{https://1inch.io/}, 2023.

\bibitem{swoopexchange}
``Swoop exchange: A decentralized exchange aggregator,'' \url{https://swoopexchange.com}, 2024.

\bibitem{matcha}
``Matcha: A decentralized exchange aggregator,'' \url{https://matcha.xyz}, 2024.

\bibitem{preguicca2018conflict}
N.~Pregui{\c{c}}a, C.~Baquero, and M.~Shapiro, ``Conflict-free replicated data types (crdts),'' \emph{arXiv preprint arXiv:1805.06358}, 2018.

\bibitem{enes2019efficient}
V.~Enes, P.~S. Almeida, C.~Baquero, and J.~Leit{\~a}o, ``Efficient synchronization of state-based crdts,'' in \emph{2019 IEEE 35th International Conference on Data Engineering (ICDE)}.\hskip 1em plus 0.5em minus 0.4em\relax IEEE, 2019, pp. 148--159.

\bibitem{google_bigquery}
``Google bigquery: Serverless, highly scalable, and cost-effective multi-cloud data warehouse,'' \url{https://cloud.google.com/bigquery}, 2024.

\bibitem{zksyncapi}
\BIBentryALTinterwordspacing
``Zksync block explorer api.'' 2021. [Online]. Available: \url{https://docs.zksync.io/zksync-era/tooling/block-explorers}
\BIBentrySTDinterwordspacing

\end{thebibliography}


\end{document}